\documentclass[runningheads]{llncs}
\usepackage[T1]{fontenc}
\usepackage{tabularx}
\usepackage{multirow}
\usepackage{amssymb}
\usepackage{todonotes}
\usepackage{subcaption}
\usepackage{filecontents}
\usepackage{pgfplots}
\usepgfplotslibrary{statistics}
\pgfplotsset{compat=1.18}
\usepackage{graphicx}
\usepackage{mdframed}
\usepackage{enumitem}
\usepackage[hidelinks]{hyperref}
\usepackage{color}

\usepackage{thmbox} 

\newcounter{mydefinition}

\begin{document}
%
\title{SciMantify - A Hybrid Approach for the Evolving Semantification of Scientific Knowledge}
\titlerunning{SciMantify - Evolving Semantification of Scientific Knowledge}
%
\author{Lena John\inst{1}\orcidID{0009-0007-2097-9761}  \and
Kheir Eddine Farfar\inst{1}\orcidID{0000-0002-0366-4596}\and
Sören Auer\inst{1}\orcidID{0000-0002-0698-2864}\and
Oliver Karras\inst{1}\orcidID{0000-0001-5336-6899}
}

\authorrunning{L. John et al.}
\institute{TIB - Leibniz Information Centre for Science and Technology, Hannover, Germany 
\email{\{lena.john, kheir.farfar, soeren.auer, oliver.karras\}@tib.eu}}
\maketitle              
\begin{abstract}
Scientific publications, primarily digitized as PDFs, remain static and unstructured, limiting the accessibility and reusability of the contained knowledge. At best, scientific knowledge from publications is provided in tabular formats, which lack semantic context. A more flexible, structured, and semantic representation is needed to make scientific knowledge understandable and processable by both humans and machines. We propose an evolution model of knowledge representation, inspired by the 5-star Linked Open Data (LOD) model, with five stages and defined criteria to guide the stepwise transition from a digital artifact, such as a PDF, to a semantic representation integrated in a knowledge graph (KG). Based on an exemplary workflow implementing the entire model, we developed a hybrid approach, called \textit{SciMantify}, leveraging tabular formats of scientific knowledge, e.g., results from secondary studies, to support its evolving semantification. In the approach, humans and machines collaborate closely by performing semantic annotation tasks (SATs) and refining the results to progressively improve the semantic representation of scientific knowledge. We implemented the approach in the Open Research Knowledge Graph (ORKG), an established platform for improving the findability, accessibility, interoperability, and reusability of scientific knowledge. A preliminary user experiment showed that the approach simplifies the preprocessing of scientific knowledge, reduces the effort for the evolving semantification, and enhances the knowledge representation through better alignment with the KG structures.

\end{abstract}

\section{Introduction}
Publications remain the primary medium for scientific communication~\cite{Johnson.2018}. Despite efforts to digitize them as PDFs, scientific knowledge largely remains static and unstructured~\cite{Bless.2023}. The next step in digital transformation calls for flexible, structured, and semantic representations to make knowledge more accessible and usable by humans and machines~\cite{Karras.2021,Karras.2023a}. Structured, machine-actionable knowledge is increasingly necessary due to the growing volume of publications and the demand for reusability~\cite{Stocker.2023}.
While some approaches, such as \mbox{\textit{SciKGTeX}}~\cite{Bless.2023}, enable \textit{FAIR-by-Design} publications~\cite{Stocker.2022}, most are only published as PDFs. At best, data is shared in tabular formats, such as results from secondary studies, which offer promising potential for semantification according to the \href{https://www.cs.ox.ac.uk/isg/challenges/sem-tab/}{\textcolor{blue}{SemTab Challenge}}~\cite{Hassanzadeh.2024}.
In this paper, we propose an evolution model of knowledge representation inspired by the \href{https://5stardata.info/en/}{\textcolor{blue}{5-star Linked Open Data (LOD) model}}. This model defines five stages with criteria to guide the transformation from static digital artifacts, e.g., PDFs, to semantic representations integrated into a knowledge graph (KG). We illustrate the implementation of the model through an exemplary workflow based on the established services \href{https://ask.orkg.org/}{\textcolor{blue}{ORKG Ask}}~\cite{Auer.2025} and \href{https://orkg.org/}{\textcolor{blue}{ORKG}}~\cite{Stocker.2023}. 
Building on this model, we introduce \textit{SciMantify}, a hybrid approach that leverages tabular formats to support the gradual semantification of scientific knowledge. Hybrid methods, combining human insight and machine automation, emerged as a promising solution to this task~\cite{Dorodnykh.2023}. In \textit{SciMantify}, humans and machines collaboratively perform semantic annotation tasks (SATs), refining results and incrementally improving semantic representations. The approach is implemented within ORKG, a core service in Germany’s National Research Data Infrastructure for FAIR scientific knowledge~\cite{Karras.2024,Stocker.2023}.
We evaluated \textit{SciMantify} in a preliminary user experiment with eight participants. Results show high usability (SUS score: 87.5)~\cite{Lewis.2018}, and participants agreed it significantly reduces preprocessing and semantification effort, averaging 17 minutes overall. We provide the following contributions: 1) The evolution model of knowledge representation, 2) The hybrid approach \textit{SciMantify}, 3) A first release of \textit{SciMantify} in the ORKG, and 4) Preliminary results indicating promising support by \textit{SciMantify}.

\section{Related Work}\label{sec:rw}
\textbf{Evolution of Knowledge Representation.} Liang et al.~\cite{Liang.2021} propose a three-layer network combining citation and content analysis to trace knowledge flow and evolution.
Li et al.~\cite{Li.2021} present the MGraph approach, a semantic data model that transforms unstructured data into structured to track concept evolution over time.
In contrast, our approach presents a generic evolution model emphasizing the transition from unstructured scientific knowledge in a digital artifact to a flexible, structured, and semantic representation integrated in a KG.

\noindent\textbf{Semantic Annotation.} 
Semantic table annotation (STA) enriches tabular data by linking it to KGs, enhancing its meaning and interoperability~\cite{Dorodnykh.2021}. Recent approaches, with significant contributions by the SemTab challenge~\cite{Hassanzadeh.2024}, include heuristic, feature engineering, and deep learning-based methods~\cite{Liu.2023}. Tools like DAGOBAH UI~\cite{Huynh.2021} and TabbyLD2-Client~\cite{Dorodnykh.2023} offer user-friendly interfaces: The former focuses on automation, while the latter supports hybrid annotation with manual refinement for broader accessibility.

STA faces key challenges, such as handling context, data heterogeneity, and metadata~\cite{Cremaschi.2024,Liu.2023}, as well as unmatched entities and incomplete KGs, that reduce annotation accuracy~\cite{Cremaschi.2024,Huynh.2021}. Most works focus on tables with single value-cells, limiting real-world application~\cite{Dorodnykh.2023,Huynh.2021,Liu.2023}, and often lack intuitive UIs~\cite{Cremaschi.2024,Dorodnykh.2023}.
To address these issues, we propose \textit{SciMantify}, a hybrid approach introducing SATs as human-machine collaboration. Unlike fully automated methods that trade off accuracy for efficiency~\cite{Liu.2023}, \textit{SciMantify} aims to achieve both through integration with the ORKG’s user-friendly UI to better support non-technical users.

\section{Evolution Model, Workflow, and Hybrid Approach}\label{sec:model,workflow,approach}

\subsection{Evolution Model of Knowledge Representation}\label{sec:evolution_model}
\begin{figure}[htb]
    \vspace{-0.5cm}
    \centering    
    \includegraphics[width=0.57\textwidth]{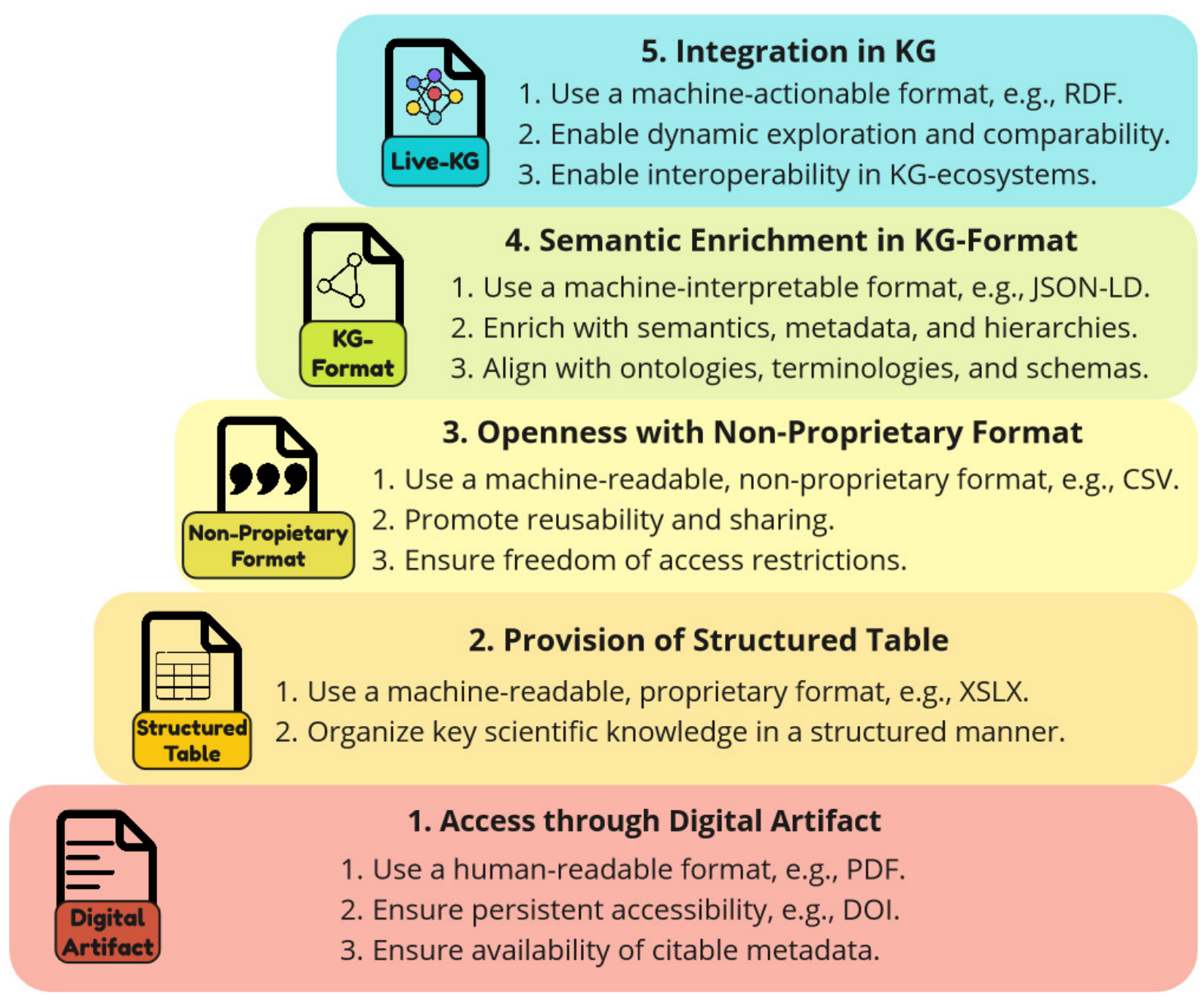}
    \caption{Evolution model of knowledge representation.}
    \label{fig:knowledge_evolution_model}
    \vspace{-0.5cm}
\end{figure}

The starting point of our work is the evolution model of knowledge representation (see~\figurename{~\ref{fig:knowledge_evolution_model}}), which we developed iteratively by testing, evaluating, and refining it through its application to different digital artifacts.

Inspired by the 5-star LOD model, our five-stage evolution model defines criteria for transitioning from unstructured scientific knowledge in a digital artifact to a semantic representation integrated in a KG. Unlike the LOD model, which focuses on the progressive openness and interoperability of data, our model focuses on the specific challenges to represent scientific knowledge.

The first stage \textbf{Access through Digital Artifact} ensures scientific knowledge is accessible in human-readable digital formats, e.g., PDFs, with stable identifiers like DOIs for long-term availability. It also emphasizes citable metadata to support reliable knowledge integration. The second stage \textbf{Provision of Structured Table}  structures key scientific knowledge in a tabular format using machine-readable formats, e.g., XLSX, enabling preprocessing, computational analysis, and efficient data management.
The third stage \textbf{Openness with Non-Proprietary Format} enhances accessibility and reusability by using open, non-proprietary formats, e.g., CSV, promoting interoperability and open science without technical or legal barriers.
The fourth stage \textbf{Semantic Enrichment in KG-Format} enriches scientific knowledge with semantics, metadata, and hierarchies in machine-interpretable formats, e.g., JSON-LD, improving interpretability, linking, and contextualization for better integration in KGs.
The fifth stage \textbf{Integration in KG} integrates scientific knowledge into KGs using machine-actionable formats, e.g., RDF or OWL, enabling dynamic exploration, comparability, and interoperability across ecosystems to support advanced data analysis and interdisciplinary collaboration.\vspace{1mm}

\subsection{Exemplary Workflow}\label{sec:workflow}
\begin{figure*}[htb]
    \vspace{-0.5cm}
    \centering
    \includegraphics[width=\textwidth]{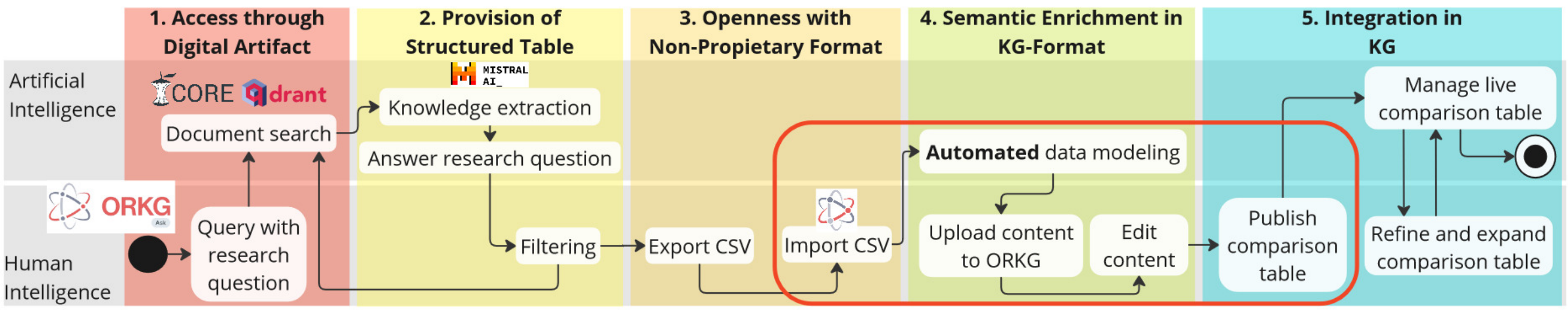}
    \caption{Exemplary workflow demonstrating the implementation of the evolution model. The red box outlines the application context of the hybrid approach.}
    \label{fig:workflow_ask_orkg}
    \vspace{-0.5cm}
\end{figure*}

The workflow combines two established services ORKG Ask~\cite{Auer.2025} and ORKG~\cite{Stocker.2023} to demonstrate the transition from unstructured scientific knowledge to semantic representations in a KG. As shown in~\figurename{~\ref{fig:workflow_ask_orkg}}, the activity diagram maps each step to a model stage, including human or machine involvement and service usage.

ORKG Ask answers natural language queries by retrieving relevant publications (Stage~1), extracting key knowledge, and generating a synthesized answer and comparison table (Stage~2), which users can refine and export as CSV (Stage~3). ORKG allows CSV import (Stage~3), applies automated data modeling to create initial semantic structures (Stage~4), and enables users to manually edit and publish the final comparison table (Stage~5).
Though based on ORKG Ask and ORKG, the workflow is service-agnostic, tools like \href{https://elicit.com/}{\textcolor{blue}{Elicit}} or
\href{https://www.wikidata.org}{\textcolor{blue}{Wikidata}} could be used instead. This workflow does not only show that the evolution model is feasible with current tools but also highlights key limitations, such as the strict separation of human and machine contributions.
As Liu et al.~\cite{Liu.2023} emphasize, modern UIs are crucial for enhancing human-machine collaboration. To address this issue, we introduce \textit{SciMantify}, a hybrid approach integrated into ORKG that supports collaborative semantification, particularly in the critical stages three to five (cf.~\figurename{~\ref{fig:workflow_ask_orkg}}, red box), where current support is lacking.

\subsection{Hybrid Approach: SciMantify}\label{sec:hybrid_approach}
\begin{figure}[htb]
    \vspace{-0.5cm}
    \centering
    \includegraphics[width=0.85\textwidth]{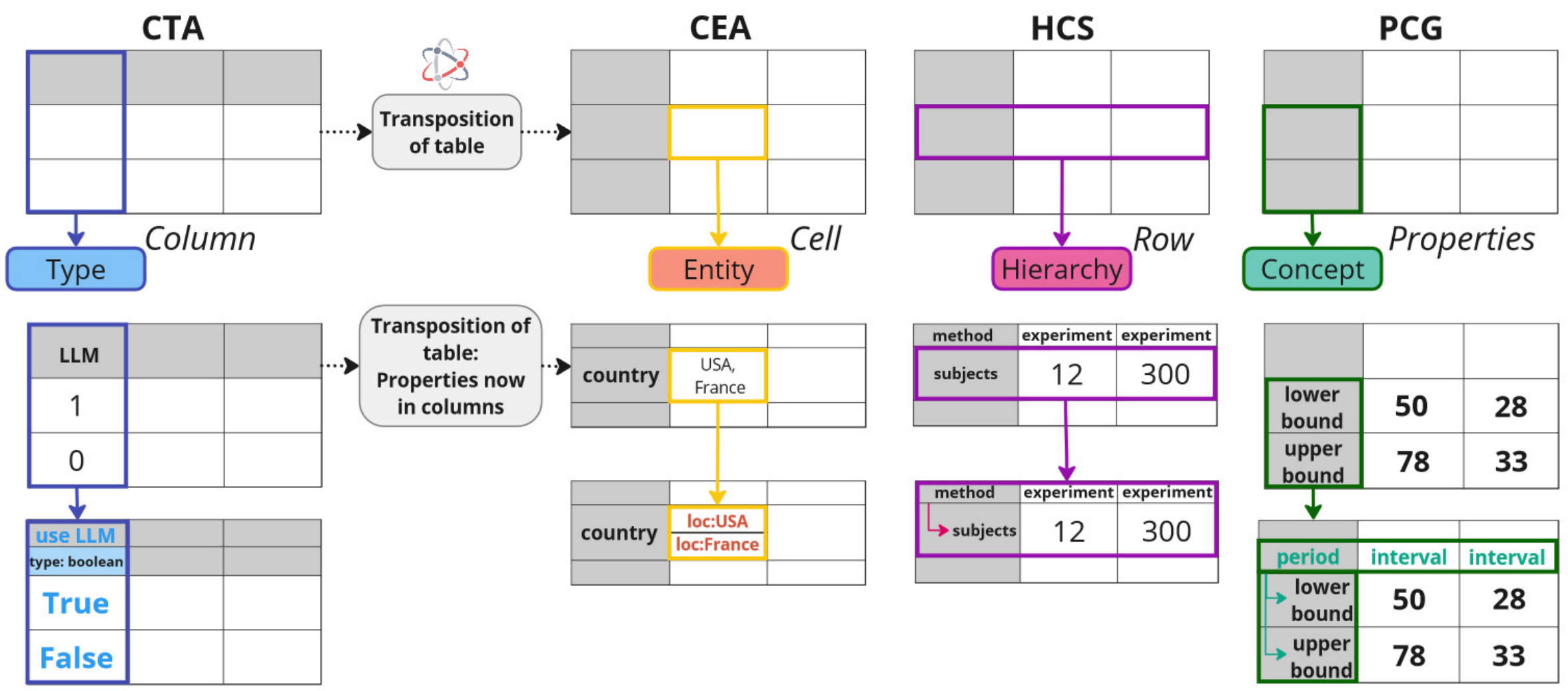}
    \caption{Semantic annotation tasks: CTA, CEA, HCS and PCG.}
    \label{fig:semantic-annotations}
    \vspace{-0.5cm}
\end{figure}

The core idea of \textit{SciMantify} is to introduce semantic annotation tasks (SATs) as collaborative efforts between humans and machines. As shown in~\figurename{~\ref{fig:use_case_annotation_tasks}}, we define four SATs that improve the automated assignment of predicates and data types during CSV import, and supporting the evolving semantification of scientific knowledge of the uploaded content in the ORKG. Inspired by the SemTab Challenge tasks~\cite{Hassanzadeh.2024}, we propose four SATs: CTA, CEA, HCS, and PCG, that enable progressive refinement (see~\figurename{~\ref{fig:semantic-annotations}}):

\begin{figure}[htb]
    \centering
    \includegraphics[width=0.7\textwidth]{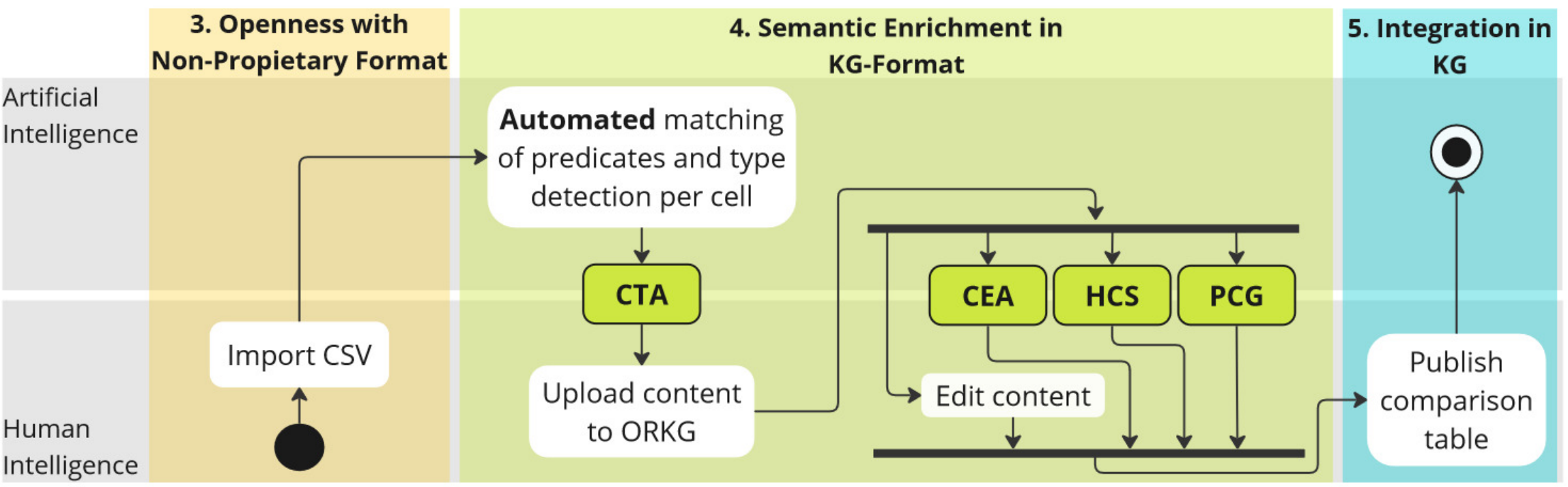}
    \caption{Hybrid approach in context of the exemplary workflow (cf. \figurename{~\ref{fig:workflow_ask_orkg}}).}
    \label{fig:use_case_annotation_tasks}
    \vspace{-0.5cm}
\end{figure}

\noindent\textbf{CTA (Column Type Annotation)} 
assigns data types and semantic properties to columns. The machine suggests types, e.g., boolean or integer, and properties; the human reviews and corrects them for contextual accuracy.\\ 
\textbf{CEA (Cell Entity Annotation)} 
links cell values to KG entities. The machine suggests matches (including handling multi-value cells by splitting), while the human verifies or refines them to improve alignment and linking.\\ 
\textbf{HCS (Hierarchical Content Structuring)}
 organizes rows into hierarchies by identifying sub-properties (rows) of existing concepts in other rows. The human defines structural relationships, and the machine applies them consistently.\\ 
\textbf{PCG (Property Concept Grouping)}
promotes reuse by grouping related properties under a new concept. The human creates the grouping, and the machine ensures consistent application across the entire table.

\section{Implementation}\label{sec:implementation}
Given the advanced state of the ORKG, we adopted an agile development approach for integrating \textit{SciMantify}. The \href{https://orkg.org/csv-import}{\textcolor{blue}{first release}} focuses on CTA and CEA tasks by embedding human input into existing automated processes. The remaining tasks will follow in the second release.

\noindent\textbf{CTA.}
The original ORKG CSV import auto-assigned predicates and treated unmatched content as strings, without editing options. Changes required modifying and re-uploading the CSV file. We enhanced this task by enabling users to edit properties and data types directly in the UI with machine support. Predicates are still auto-assigned, but users can revise them. Data types are inferred via majority voting, and inconsistencies are flagged in real-time for resolution.

\noindent\textbf{CEA.}
The original contribution editor allowed only basic edits. While users could add new cell values (suggested by the machine), aligning existing values with ORKG entities required manual deletion and re-entry, offering no true CEA support. For this reason, we introduced the ``Semantify'' function: Users can now align existing values with suggested entities or create new ones. Enumerations in a cell can be split and semantically aligned to enhance data integration.

\section{Evaluation}\label{sec:evaluation}
To assess the hybrid approach proposed in this paper, we conducted a preliminary user experiment focusing on the first release of \textit{SciMantify}, which integrates humans into automated processes for CTA and CEA. The aim was to explore the approach’s effectiveness, efficiency, and user satisfaction in supporting the evolving semantification of scientific knowledge in the ORKG.

\vspace{-0.1cm}
\begin{mdframed} \textit{Research question:} How does the implementation of the hybrid approach (CTA \& CEA) support users with prior knowledge of semantic data modeling in terms of effectiveness, efficiency, and satisfaction in the ORKG? \end{mdframed}
\vspace{-0.1cm}

Participants performed the CTA and CEA tasks using a simplified ORKG comparison table~\cite{Lozynska.2024}, adapted to emphasize data modeling. Eight participants familiar with the ORKG took part, half of whom regularly used the CSV import.

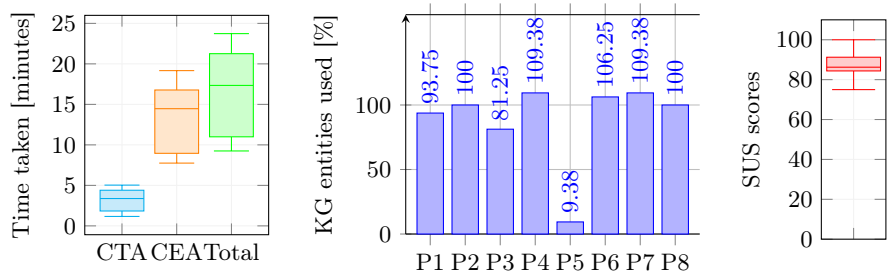
\begin{figure}[!ht]
    \centering
    \begin{subfigure}[b]{0.33\textwidth}
        \centering
        \raisebox{0.04\height}{
        \begin{tikzpicture}
            \begin{axis}[
                width=\linewidth,
                height=4.5cm,
                boxplot/draw direction=y,
                ylabel={Time taken [minutes]},
                xtick={1,2,3},  
                ytick={0, 300, 600, 900, 1200, 1500},
                grid=both,
                grid style={line width=.1pt, draw=gray!10},
                xticklabels={\shortstack{CTA}, \shortstack{CEA}, Total},
                yticklabel=\pgfmathparse{int(\tick/60)}\pgfmathresult,
            ]
            \addplot+[boxplot,fill=cyan!20,draw=cyan] table[row sep=\\, y index=0] {
                data \\
                256 \\
                264 \\
                81 \\
                302 \\
                70 \\
                120 \\
                149 \\
                264 \\
            };
    
            \addplot+[boxplot,fill=orange!20, draw=orange] table[row sep=\\, y index=0] {
                data \\
                1150 \\
                968 \\
                768 \\
                1122 \\
                485 \\
                555 \\
                465 \\
                968 \\
            };
    
            \addplot+[boxplot, fill=green!20, draw=green] table[row sep=\\, y index=0]{
                data \\
                1406\\
                1232\\
                849\\
                1424\\
                555\\
                675\\
                614\\
                1232\\
            };
        \end{axis}
        \end{tikzpicture} 
        }
        \caption{Times for CTA task, CEA task, and in total.}
        \label{fig:times_taken}
    \end{subfigure}
    \hfill
    \begin{subfigure}[b]{0.45\textwidth}
        \centering 
        \begin{tikzpicture}
            \begin{axis}[
                width=\linewidth,
                height=4.5cm,
                ybar, 
                ymin=0, ymax=170, 
                ytick={0,50,100},
                ylabel={KG entities used [\%]},
                symbolic x coords={P1, P2, P3, P4, P5, P6, P7, P8}, 
                xtick=data, 
                nodes near coords,
                every node near coord/.append style={rotate=90, anchor=west},
                enlargelimits=true,
                axis y line=left, 
                grid=both, 
            ]
            \addplot coordinates {
                (P1,93.75) (P2,100) (P3,81.25) (P4,109.375) 
                (P5,9.375) (P6,106.25) (P7,109.375) (P8,100)
            };
            
            \end{axis}
        \end{tikzpicture}
        \caption{Ratio of entities used and 32 entities of the original comparison table.}
        \label{fig:barplot_resources}
    \end{subfigure}
    \hfill
    \begin{subfigure}[b]{0.2\textwidth}
        \centering
        \raisebox{0.095\height}{
        \begin{tikzpicture}
            \begin{axis}[
                width=\linewidth,
                height=4.5cm,
                boxplot/draw direction=y,
                ylabel={SUS scores},
                xtick={1},  
                ytick={0,20,...,100},
                ymin=0,
                grid=both,
                grid style={line width=.1pt, draw=gray!10},
                xticklabels={},
            ]
            \addplot+[boxplot,fill=red!20,draw=red] table[row sep=\\, y index=0] {
                data \\
                85 \\
                90\\
                87.5\\
                100\\
                75\\
                95\\
                85\\
                82.5\\
            };
        \end{axis}
        \end{tikzpicture} }
        \caption{SUS scores for \textit{SciMantify}.}
        \label{fig:sus_boxplot}
    \end{subfigure}
    \caption{Objective assessment of effectiveness, efficiency, and satisfaction.}
    \label{fig:objective_plots}
    \vspace{-0.5cm}
\end{figure}

Subjective feedback indicated that most participants found \textit{SciMantify} effective and efficient, particularly for CTA and CEA. However, the predicate selection feature received mixed feedback, with some participants overlooking it. Task completion times (see \figurename{~\ref{fig:times_taken}}) averaged 16:38 $\pm$ 6:04, [9:15 - 23:44] minutes, with 3:08 $\pm$ 1:33, [1:10 - 5:02] for CTA and 13:30 $\pm$ 4:41, [7:45 - 19:10] for CEA. The number of reused KG entities (see \figurename{~\ref{fig:barplot_resources}}) varied, with most participants aligning with or exceeding the original table, except one, who was unfamiliar with the contribution editor. The overall usability (see \figurename{~\ref{fig:sus_boxplot}}) was rated high, with a mean SUS score of 87.5 $\pm$ 7.68, [75 - 100], i.e., excellent usability~\cite{Lewis.2018}.

In summary, the experiment showed encouraging initial results for the hybrid approach. However, future larger studies are need to compare \textit{SciMantify} with a baseline of the original ORKG workflow to quantify improvements.

\section{Discussion}\label{sec:discussion}
A more structured and semantic representation of scientific knowledge is essential to advancing the digital transformation of scientific communication. To address this, we proposed an evolution model of knowledge representation, demonstrated through an exemplary workflow and implemented as \textit{SciMantify}, a hybrid approach supporting human-machine collaboration in the semantification process.
Our preliminary user experiment indicates that \textit{SciMantify} effectively supports the transformation of scientific content from tabular formats to semantic representations. With an average completion time of 17 minutes (3 for CTA, 14 for CEA), results show efficient preprocessing and highlight the need for human input in refining machine-suggested entity alignments. The high usability score (SUS 87.5) confirms the system's user-friendliness and overall satisfaction.
The integration of CTA and CEA tasks proved essential, though feedback also pointed to areas for improvement. In particular, predicate selection was often overlooked, suggesting the need for better UI guidance. Additionally, variability in CEA task times indicates that users would benefit from more structured support when aligning entities, especially in complex contexts. Several participants also expressed the need for hierarchical structuring and grouping, aligning with our planned integration of HCS and PCG in the next release of \textit{SciMantify}.
While results are promising, the evaluation has limitations. The study included only eight participants, and the scenario was based on a simplified, fictitious example. Broader studies using real-world data and a baseline comparison are needed to validate the findings. Furthermore, only CTA and CEA tasks were implemented so far. The full potential of the hybrid approach will be assessed after integrating HCS and PCG in the upcoming release.

\section{Conclusion and Future Work}\label{sec:conclusion}
The proposed evolution model for knowledge representation offers a structured approach to transforming static, unstructured knowledge into flexible, semantic formats. Based on the exemplary workflow that demonstrates the implementation of the evolution model, we developed \textit{SciMantify}, a hybrid approach combining human expertise with machine automation to evolve the representation of scientific knowledge from tabular formats to semantic representations in a KG.
A preliminary experiment within the ORKG context showed high usability and positive feedback, highlighting \textit{SciMantify}’s ability to reduce preprocessing and semantification efforts. Future work includes implementing the remaining semantic annotation tasks HCS and PCG, and adding AI-driven suggestions to better support complex content. We plan a broader user study to assess performance across research domains and compare it to a baseline workflow to increase generalizability. Overall, the evolution model and \textit{SciMantify} represent a key step toward more accessible, reusable, and interoperable scientific knowledge.

\bibliographystyle{splncs04}
\bibliography{references-nodois}

\end{document}